\documentclass[aps,twocolumn,showpacs,prl]{revtex4-1}
\usepackage[english]{babel}
\usepackage{amsmath,amssymb} 
\usepackage{times}
\usepackage{epsfig} 

\begin{document} 

\title{Dielectric breakdown in spin polarized Mott insulator}

\author{Zala Lenar\v ci\v c$^{1}$ and Peter Prelov\v sek$^{1,2}$}
\affiliation{$^1$J.\ Stefan Institute, SI-1000 Ljubljana, Slovenia}
\affiliation{$^2$Faculty of Mathematics and Physics, University of
Ljubljana, SI-1000 Ljubljana, Slovenia}
\begin{abstract}
Nonlinear response of a Mott insulator to external electric field, 
corresponding to dielectric breakdown phenomenon, is studied within
of a one-dimensional half-filled Hubbard model.  
It is shown that in the limit of nearly spin polarized insulator the decay rate 
of the ground state into excited holon-doublon pairs can be evaluated 
numerically as well to high 
accuracy analytically. Results show that the threshold field depends 
on the charge gap  as $F_{th} \propto \Delta^{3/2}$. 
Numerical results on small systems indicate  on the persistence of  a similar 
mechanism for the breakdown for  decreasing magnetization 
down to unpolarised system.  
\end{abstract}


\maketitle
The nonlinear response to external fields and more general nonequilibrium properties 
of strongly correlated electrons and Mott insulators in particular \cite{imad} 
are getting more attention in recent years, also
in connection with powerful novel experimental techniques, e.g. the
pump-probe experiments  on Mott insulators \cite{pp}, as well as novel systems, 
the prominent example being the driven ultracold atoms within the insulating phase  
 \cite{stroh}. In this connection, one of the basic phenomena to be understood 
is the dielectric breakdown in Mott insulators, studied experimentally in 
effectively one-dimensional (1D) systems more than a decade ago \cite{tagu}.
The concept of Landau-Zener (LZ)  single-electron tunneling  \cite{zen1,land} 
as a standard approach  to dielectric  breakdown of band insulators  \cite{zen2} 
is not straightforward to generalize to correlated electrons \cite{oka1,oka2,oka3}.
Theoretical efforts have been so far restricted to the prototype Hubbard
model at half-filling. In 1D  numerical approaches have given 
some support to analytical approximations for the most interesting quantity
being the threshold field $F_{th}$ and its dependence on the charge gap $\Delta$ 
\cite{oka2}, typically revealing a LZ type dependence $F_{th} \propto
\Delta^2$. Different dependence is found numerically within the 
dynamical-mean-field-theory approach \cite{ecks} as relevant for high dimensions 
$D \gg 1$.   

In this Letter we approach the problem of a dielectric breakdown from 
a partially spin polarized Mott insulator.  We use the fact that the ground state (g.s.) 
of the 1D Hubbard model is insulating at any spin polarization with the charge gap
modestly dependent on the magnetization $m$. In particular, a single 
spin excitation in fully polarized system $m\sim 1/2$, i.e. $\Delta S=1$ state, 
can be studied exactly numerically as well as to 
high accuracy analytically. The relevant mechanism
for the decay of  the g.s. under constant external field $F$ is the creation of
holon-doublon (HD) pairs. We show that due to the dispersion-less g.s. the similarity to
the LZ tunneling is only partial and leads to a different scaling $F_{th} 
\propto \Delta^{3/2}$. Furtheron we study numerically on small systems 
also the model with $\Delta S>1$, $m<1/2$ in a finite field $F$. 
Results indicate that the decay mechanism remains qualitatively and 
even quantitatively similar at polarizations $m<1/2$, 
in particular for larger $\Delta$ whereby the most interesting case is clearly the 
unpolarized  $m=0$ system.

In the following we study the prototype 1D Hubbard model,
\begin{equation}
H=- t \sum_{i \sigma} (e^{ i \phi} c^\dagger_{i+1,\sigma} c_{i \sigma} + 
\mathrm{H.c}) + U \sum_i n_{i\uparrow} n_{i\downarrow}, \label{ham}
\end{equation}
with periodic boundary conditions (p.b.c.) 
where $c^\dagger_{i\sigma},c_{i\sigma}$ are creation (annihilation) operators for
electrons at site $i$ and spin $\sigma=\uparrow,\downarrow$. 
The action of an external electric field $F$  is induced via the  
Peierls phase $\phi$ (vector potential) and its time dependence, i.e. 
$\dot \phi( \tau)= e_0 F( \tau)a_0 /\hbar$. Furtheron we use units 
$\hbar=e_0=a_0=1$, as well as we put $t=1$ defining the unit of energy. 
In such a model we investigate finite systems 
of length $L$  and at half-filling 
$N_u+ N_d=L$ but in general at finite total spin,
$S^z = (N_u- N_d)/2$ and magnetization 
$m=S^z/L$.

Let us first consider the problem of a single overturned spin, i.e. 
$\Delta S^z=L/2-S^z=1$. Here, basis wavefunctions $|\varphi_{jm}\rangle$
correspond to an empty site (holon) at site $j$ and a doubly occupied site 
(doublon) at site $m$. Taking into account the translational symmetry of the model (\ref{ham})
with p.b.c. (even with time dependent $\phi(\tau)$) at given (total) momentum $q=2\pi m_q/L$ 
the relevant basis is 
$ |\Psi_{q}^{l}\rangle=(1/\sqrt{L})\sum_{j}e^{iqj}|\varphi_{j,j+l}\rangle , l \in [0,L-1]$.
At fixed $\phi$ adiabatic eigenfunctions  can be then searched in the form 
$|\psi\rangle=\sum_{j}d_{j}|\Psi_{q}^{j}\rangle$  leading to the eigenvalue equation, 
\begin{equation}
 -\frac{1}{U}=\frac{1}{L}\sum_{q'}\frac{1}{E-U + 2(\cos(q'-\phi) + \cos(q'-\phi-q))}
 \label{eeq}.
\end{equation}
In the limit $L \to \infty$ the g.s. energy $E_0$ representing the holon-doublon (HD) bound state
can be expressed explicitly as $E_{0}=U - (U^2 + 16 \cos^2 (q/2) )^{1/2}$. We note
that (in spite of the $q$-dependence) g.s. states for all $q$ are nonconducting
since from Eq.~(\ref{eeq} )
it follows that the charge stiffness ${\cal D}_0 \propto  \partial^2 E_{0}/\partial \phi^2 \to 0 $ 
for $L \to \infty$. On the other hand, excited states form a continuum with lower edge at
$E_{1} = U-4 \cos(q/2)$. 

Since $\phi(\tau)$ conserves total $q$ we furtheron consider 
only solutions within the $q=0$ subspace representing the absolute g.s. wavefunction
$|0\rangle$ with $d^0_j= A e^{-\kappa |j|} e^{i\phi j}$ and $A=\sqrt{\tanh \kappa}$.  Here,
the charge gap $\Delta=E_1-E_0$ and the related g.s. localization parameter 
$\kappa$ are given by
\begin{equation}
\Delta=-4+\sqrt{U^2 + 16}  = 4(\cosh\kappa -1). \label{delta}
\end{equation}
When we consider the time-dependent $\phi(\tau)$ we have to deal
at finite $L$ with adiabatic states $E_n(\phi)$ as, e.g., shown in Fig.~1 for  finite $L$. 
At finite $L\gg 1/\kappa$ $E_0$ is essentially  $\phi$-independent 
but the same holds as well for lowest excited states $E_n, n \gtrsim 1$ which makes an 
usual application of two-level LZ approach not straightforward to apply.  
If we suppose that due to field $F>0$ the transition probability between 
neighbouring states is high (neglecting finite size gaps between them)
 excited states are well represented by 'free' HD pair states with 
\begin{equation}
d^k_j = \frac{1}{\sqrt{L}} e^{ikj}, \quad \epsilon_k=U-4 \cos(\phi-k), \quad
k=\frac{2 \pi}{L} m_k.
\end{equation} 
As shown further relevant transitions due to time-dependent 
$\phi(\tau)$ happen to effective states $|k \rangle$ with $|m_k| \gg 1$ 
since the g.s.  $|0 \rangle$  is well localized.  

\begin{figure}[ht]
\includegraphics[width=0.43\textwidth]{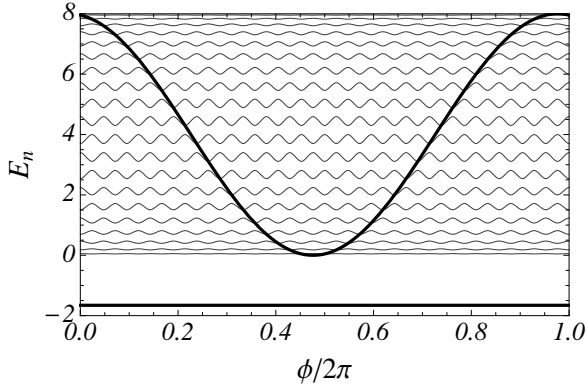}
\caption{Energy levels $E_n$ (in units of $t$) vs. phase $\phi$ for holon-doublon pair 
states in the system with $L=21$ sites and  $U=4$. Thick line represents the 
g.s. and the effective HD pair state dispersion.}
\label{fig1}
\end{figure}

Let us now consider the decay of the g.s. $|0 \rangle$ after switching 
constant field $F (\tau>0)=F,  \phi=F\tau$. We present an analysis 
for the initial decay where 
most weight is still within the g.s., i.e. $|a_0(\tau)| \gg |a_{n\neq0}(\tau)|$. In such case the excited state
 amplitude time-dependence $a_n(\tau)$  is given by 
\begin{equation}
 a_{n}(\tau)=-F \int_{0}^{\tau} d\tau' \Phi_n(\tau')   
 \exp (i\int_{0}^{\tau'}\omega_{n}(\tau'')d\tau''), \label{an0t}
 \end{equation}
where  $\Phi_n= \langle n| \partial /\partial \phi |0\rangle$ and   $\omega_{n}(\tau)=E_n(\phi)-E_0$. 

Analytically progress can be made by using effective HD states $|k\rangle$ 
as approximate excited states  with $\omega_{k}(\xi)= \epsilon_k - E_{0}=4(\cosh \kappa - \cos\xi), 
\xi=F\tau-k$. By using the relation
\begin{equation}
\langle k | 0\rangle \omega_k= \langle k | H_{0} +U-H| 0\rangle =
U \langle k | n_{0\downarrow} | 0\rangle = U A/\sqrt{L},
\end{equation}
where $H_0$ denotes only kinetic term in Eq.~(\ref{ham}),
one can express $\Phi_k$ in Eq.~(\ref{an0t}) as
\begin{equation}
\Phi_k = \langle k | \frac{\partial }{\partial \phi}  | 0\rangle = \frac{\partial}
 {\partial \phi} \langle k | 0\rangle= \frac{UA }{\sqrt{L}} \frac{\partial \omega_{k}^{-1}}{\partial \phi}.
  \label{phik}
 \end{equation}
Here, we can already realize some essential differences to 
the usual concept of of LZ tunneling, i.e., $\Phi_k$ 
and Eq.~(\ref{an0t}) do not favor transitions to lowest lying excited state but rather 
to $k \sim\kappa/\sqrt{3}$, hence the reduction to a two-level problem is not
appropriate. 

The rate of $a_k(\tau)$ following from Eqs.~(\ref{an0t}), (\ref{phik}) is not steady. 
Since we are interested in low $F$ we average it over the Bloch period 
$\tau_B=2\pi/F$ to get $\bar a= a_k(\tau_B)$ which is approximately
the same for majority of  $k$ (fixing here $k = \pi$),
\begin{eqnarray}
\bar a&=-& \frac{A U}{\sqrt L}\int_{-\pi}^{\pi}d\xi 
 \left(\frac{1}{\omega_{\pi}(\xi) } \right)^\prime 
 \exp\left (\frac{i}{F}\int_{-\pi}^\xi d\xi^\prime \omega_{\pi}(\xi') \right) \label {aav0}  \\
& \sim &\frac{i A U}{F \sqrt L}\int_{-\pi}^{\pi}d\xi 
 \exp\left (\frac{i}{F}\int_{-\pi}^\xi d\xi^\prime \omega_{\pi}(\xi') \right), \label{aav}
\end{eqnarray}  
after per partes integration of Eq.~(\ref{aav0}) and neglecting the first fast oscillating 
part, smaller also due to an additional prefactor $F$.
Final simplification for small F can be made by replacing  
$\cosh \kappa - \cos\xi \sim \xi^2/2 +
\Delta/4 $ and consequently extending integrations in Eq.~(\ref{aav}) to 
$\xi =\pm \infty$. This leads to an analytical expression for the decay 
rate $\Gamma$, defined by $|a_0|^2 \sim  \exp(-\Gamma \tau$) where $\Gamma =
L |\bar a|^2 /\tau_B$,
\begin{equation}
\Gamma=\frac{\Delta^{3/2}B(\Delta)}{3\pi F}K^2_{\frac{1}{3}}\left(\frac{\sqrt{2} \Delta^{3/2}}{3F}\right) 
\sim \frac{B(\Delta)}{\sqrt{8}}\exp\left(-\frac{(2\Delta)^{3/2}}{3F}\right) \label{gamma}
\end{equation}
where $K_{1/3}(x)$ is the modified Bessel function and 
$B(\Delta)=\Delta(\Delta+8)^{3/2}/(\Delta+4)$, and the last  exponential
approximation is valid for small enough $\Gamma$.
The main conclusion of the analysis is that  $\Gamma$ in Eq.~(\ref{gamma}) 
depends on $\Delta^{3/2}/F$  unlike usual LZ theory applications
\cite{oka1,oka2} yielding $\Delta^2/F$. As the threshold field is usually defined
with the expression $\Gamma \propto \exp(-\pi F_{th}/F)$, Eq.~(\ref{gamma}) directly 
leads $F_{th} = (2\Delta)^{3/2}/(3\pi)$.

It is straightforward to verify the validity of approximations for $N_d=1$ 
via a direct numerical solution of the time-dependent Schr\"odinger equation (TDSE) 
with $\phi=F \tau$ within the full basis at $q=0$ and finite but large $L> 100$. Time
dependence of the g.s. weight $|a_0(\tau)|^2$ is presented in
Fig.~2  for typical case $U=4$ and different fields $F=0.2-0.5$.
Results for the case of an instantaneous switching $F(\tau>0)=F$
(shown for $F=0.5$) reveal some oscillations (with the frequency proportional to the gap
$\Delta$) but otherwise clear exponential decay 
with well  defined $\Gamma$. In order to minimize the fast-switching effect
we use in Fig.~2 and furtheron mostly smooth transient \cite{ecks}, 
i.e.,  field increases as $F(\tau<0) = F \exp( 3\tau/\tau_B)$ to its final 
value $F(\tau>0)=F$. 

\begin{figure}[ht]
\includegraphics[width=0.43\textwidth]{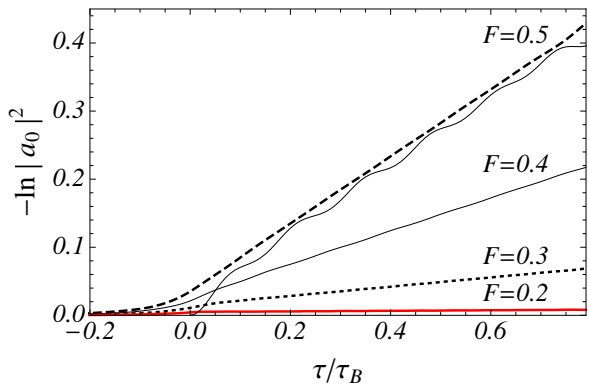}
\caption{(Color online)  a) Ground state  weight $\ln|a_0|^2$ vs. time 
$\tau/\tau_B$ for  $U=4$ and different fields $F=0.2 - 0.5$. For $F=0.5$ the 
comparison of results for smoothly and instantaneously switched $F(\tau)$ is presented 
while for $F<0.5 $ only smooth switching is used. }
\label{fig2}
\end{figure}

In Fig.~3 we compare results for $\Gamma$ as obtained 
via three different methods: a) direct numerical
solution of TDSE, b) analytical approximation with an average decay rate 
into free HD states, numerically integrating Eq.~(\ref{aav0}), and c) the explicit expression (\ref{gamma})
where additional simplification of the parabolic dispersion of excited states is used. 
The agreement between different methods is satisfactory
essentially within the whole regime of small $\Gamma$ and deviations between
analytical and numerical results become visible only for large $\Gamma \sim 0.1$.
Moreover, results confirm the expected variation $\ln\Gamma \propto 1/F$ essentially
in the whole investigated range of  $F$.

\begin{figure}[ht]
\includegraphics[width=0.43\textwidth]{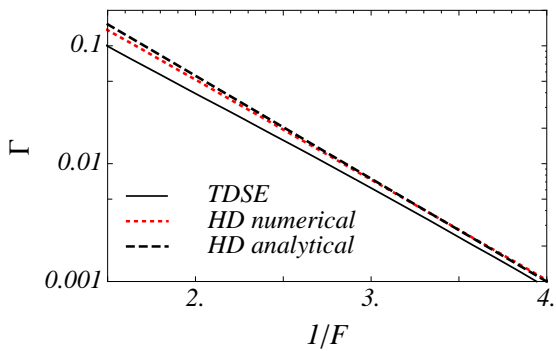}
\caption{(Color online) Ground state decay rate $\Gamma$ (log scale) 
vs. $1/F$ for $U=4$ as evaluated by direct numerical solution of TDSE (full line), decay into free HD states,
numerically integrating Eq.~(\ref{aav0}) (dotted line), and analytical expression, Eq.~(\ref{gamma}) (dashed line).    }
\label{fig3}
\end{figure}

One can assume that a similar mechanism of the dielectric breakdown via the
decay into free HD pairs remains valid at finite deviations $N_d >1$ and 
$m<1/2$. In order to test this scenario we perform the numerical solution of 
TDSE for the model, Eq.~(\ref{ham}), with the finite field $F(\tau)$. 
Calculation for all $S^z$ sectors 
covering the whole regime $0\le m<1/2$  are performed on finite Hubbard chains with 
up to $L=16$ sites using the Lanczos procedure both for the determination
of the initial g.s. wavefunction $|0 \rangle $ as well as for the time 
integration of the TDSE \cite{pb} within the full basis for 
given quantum numbers $N_d, N_u, q=0$ reaching up to $N_{st} \sim 10^7$ basis
states. We use everywhere smooth transient for the field $F(\tau)$.
Since the decay rate of the g.s. weight $|a_0|^2$  is expected to scale 
with the number of overturned spins $N_d$  the relevant 
quantity to follow and compare is $(1/N_d) \ln |a_0|^2(\tau)$.

In Figs.~4,5 we present numerical results for time dependence of
normalized g.s. weight $\ln |a_0|^2/ N_d$ as obtained via a
direct solution of the TDSE for $L=16$ with the whole range
of magnetization $1/2 > m \geq 0$  (relevant $1 \leq N_d \leq L/2$)
 for two cases of $U=4,10$,
respectively, and the span of appropriate fields $F$.  Examples
are chosen such to represent  charge gap (for 
a single HD pair) being small $\Delta \sim 1.3 <W$
and large $\Delta \sim 6.5 > W$, respectively,  relative to the 
noninteracting bandwidth $W=4$. 

\begin{figure}[ht]
\includegraphics[width=0.43\textwidth]{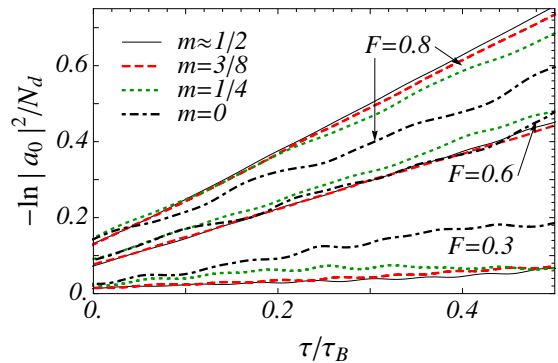}
\caption{(Color online) Normalized g.s. weight $(1/N_d) \ln |a_0|^2$
vs. time $\tau/\tau_B $ for $U =4$ and fields  $F= 0.3, 0.6, 0.8$, 
for various spin states $1\leq N_d\leq L/2$.  }
\label{fig4}
\end{figure}

\begin{figure}[ht]
\includegraphics[width=0.43\textwidth]{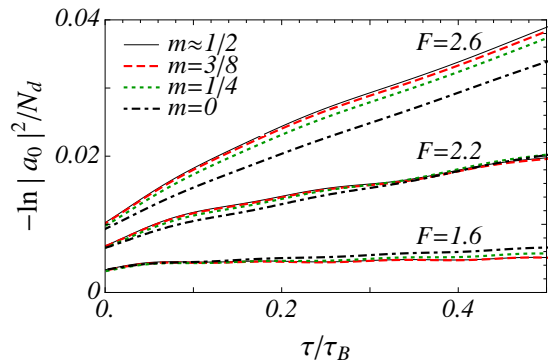}
\caption{(Color online)  The same as in Fig.~4 for $U = 10$ and $F=1.6, 2.2, 2.6$. }
\label{fig5}
\end{figure}

The main conclusion following from Figs.~4,5 is that the g.s. weight $|a_0|^2$ indeed
decays proportional to $N_d$ confirming the basic mechanism of the field-induced
creation of (nearly independent) HD pairs.  The decay rate $\Gamma$ defined 
as $|a_0|^2 \propto \exp (-\Gamma N_d \tau)$ is only moderately dependent on
$N_d$ and $m$. Results confirm that $\Gamma$ is 
essentially independent of $N_d$ in well polarized systems with $m \geq 1/4$,
which is compatible with independent decay into low concentration of HD pairs.
For larger $U=10$  in Fig.~5 the invariance of $\Gamma$ extends even to 
unpolarized situation $m=0$ ($N_d/L = 1/2$) for intermediate fields 
$F \geq 2.2$.  

There are some visible deviations at $m \leq 1/4$
for weakest fields both in Fig.~5 for $F=1.6$ and even more for smaller $U=4$
and $F=0.3$ in Fig.~4, indicating on larger $\Gamma$ and correspondingly 
faster decay of unpolarized g.s. with $m=0$ relative to nearly saturated 
$m\sim 1/2$. Part of this enhancement of $\Gamma$ can be attributed to
the dependence of the charge gap on the magnetization $\Delta(m)$. 
The thermodynamic ($L \to \infty$) value $\Delta_0=\Delta(m=0)$ is known 
via the Bethe Ansatz solution
given by the equation  $\Delta_0=(16/U)\int_1^\infty  dx \sqrt{x^2-1}/ \sinh (2\pi x/U)$ 
\cite{lieb,ovchinnikov}.  Values
for $\Delta(m\sim 1/2)$ as given by Eq.~(\ref{delta})
are somewhat larger than $\Delta_0$ with the relative difference becoming 
more pronounced for $U <4$. Still taking into account actual $\Delta(m)$ 
some enhancement
seems to remain at $m \sim 0$ at least for weaker fields $F$ and smaller $U$. This
could indicate that the decay into HD pairs are not independent processes but
correlations due to finite concentration of $N_d/L$ enhance decay. 

Finally let us consider the threshold field for the decay $F_{th}$ 
as defined again by $\Gamma \propto \exp(-\pi F_{th}/F)$. We present results in Fig.~6
for $F_{th}$  as function of the gap $\Delta$.  To extract $F_{th}$ vs. $\Delta$ we use
numerical data for $\Gamma(F)$ obtained from numerical $|a_0|^2(\tau)$
as, e.g., shown in Figs.~2,4,5. For the reference charge gap $\Delta(m)$ we use 
 for $m\sim 1/2$ and  $m=1/4$  Eq.~(\ref{delta}), while for $m=0$ we
use exact $\Delta_0$. Some deviation between $m\sim 1/2$ and $m=1/4$ results can be 
still attributed to actually slightly smaller gap for the latter magnetization. For comparison we plot also 
the analytical result emerging from Eq.~(\ref{gamma}), 
$F_{th} \propto \Delta^{3/2}$, as well as the dependence following 
from the LZ approach \cite{oka2} with $F_{th} =\Delta^{2}/8$. 
From Fig.~6 we conclude that  the general trend $F_{th}(\Delta)$ is 
quite well represented by the single
HD pair result which deviates significantly from the LZ dependence at least for
larger $\Delta>6$. At the same time, we should note that  our numerical results
in the range $1<\Delta<2.1$ agree also well with data analyzing 
numerically the g.s. decay using the  t-DMRG method (at $m=0$) for the 
same model but bigger $L \sim 50$ \cite{oka2}.

\begin{figure}
\includegraphics[width=0.43\textwidth]{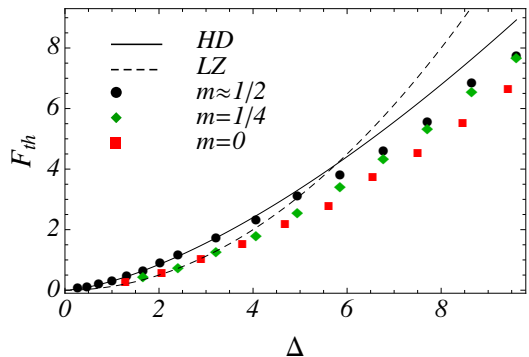}
\caption{(Color online)  Threshold field $F_{th}$ vs. charge gap $\Delta$ for 
different magnetizations $m\sim 1/2$ (given by $N_d=1$) and $m=1/4, 0$
as obtained numerically for $L=16$. Full curve (HD) represent the analytical 
approximation, Eq.~(\ref{gamma}), while the dashed curve is the LZ 
approach result from  Ref.~\cite{oka2}. }
\label{fig6}
\end{figure}

In conclusion, we have presented an analysis of the dielectric breakdown within
the Mott-Hubbard insulator starting from a spin polarized ground state. Such an 
approach has clearly an advantage since the problem can be solved up to desired
accuracy numerically but as well captured analytically. As such the 
situation can serve at least as well controlled test for more demanding situations
of  an arbitrary magnetization, in particular of an unpolarized g.s. \cite{oka3,ecks}.

The case of a nearly polarized state $N_d=1$ describes the mechanism of
the field-induced decay of the g.s. into single HD pair. Here one can follow 
differences to usual LZ-type approaches: a) the g.s. is
localized and dispersionless  within the insulator, b) the transition is not between two isolated
levels but rather to a continuum, moreover it follows from Eqs.~(\ref{an0t}),(\ref{aav0})
that matrix elements do not favor transitions to lowest excited states, c)  instead of exact
excites states, one can well use effective free HD states, d) dispersion of effective HD states 
is unlike in LZ applications not hyperbolic, e.g., $ \omega_k  \propto (k^2+\kappa^2)^{1/2}$  
but rather parabolic $\omega_k = k^2 + \kappa^2$ which is presumably the main origin
for qualitatively different behavior of the threshold field $F_{th} \propto \Delta^{3/2}$
which is a final manifestation of the distinction to usual LZ applications. On the other
hand there are some similarities. In particular the analytical expression for the average 
transition rate, Eq.~(\ref{aav}), where matrix element is integrated out, appears
analogous to two-level problem and ready for phase-integral transformation into 
imaginary plane as used originally by Landau \cite{land} and then generalized 
\cite{dykh,davi} and applied as well to breakdown problem \cite{oka3,oka4}. 
Still it is straightforward to verify that for the levels under consideration 
$\omega_k$ do not satisfy criteria for its application, but the analogy rather 
emerges through the application of the steepest descent approximation to 
Eqs.~(\ref{aav}).

The picture of the decay of the driven Mott insulator into HD
pairs remains attractive for magnetization approaching
the unpolarized g.s. There seem to be two characteristic length scales 
controlling the mechanism, the HD pair localization length $\zeta = 1/\kappa$ and the 
Stark (Bloch) localization scale $L_S=8/F$. Our results indicate that
for larger $\Delta$ (small $\zeta$) and well localized HD pairs the mechanism
of decay into nearly independent HD pairs remains at least qualitatively valid. 
On the other hand, we find indications that  for smaller $\Delta$ and weaker $F$ 
(larger $L_S$),  the decay is enhanced, i.e., pointing into the direction
of more collective driven excitations favored also in the interpretation
of experiments \cite{tagu}. It should be as well pointed out that the phenomenon 
of HD pair generation is not particularly specific to 1D systems discussed here 
but  can generalised to higher dimensional Mott insulators as well.

This work has been supported by the Program P1-0044 and the project J1-4244 
of the Slovenian  Research Agency (ARRS).


\begin{thebibliography}{17}
\bibitem{imad} M. Imada, A. Fujimori, and Y. Tokura, Rev. Mod. Phys.
{\bf 70}, 1039 (1998).
\bibitem{pp} H. Okamoto, H. Matsuzaki, T. Wakabayashi, Y. Takahashi,
and T. Hasegawa, Phys. Rev. Lett. {\bf 98}, 037401  (2007).
\bibitem{stroh} N. Strohmaier {\it et al.}, Phys. Rev. Lett. {\bf 104}, 080401  (2010).
\bibitem{tagu} Y. Taguchi, T. Matsumoto, and Y. Tokura, Phys. Rev. B {\bf 62}, 7015 (2000).
\bibitem{land} L. Landau, Sov. Phys. {\bf 1}, 89 (1932). 
\bibitem{zen1} C. Zener, Proc. R.Soc. A {\bf 137}, 696 (1932). 
\bibitem{zen2} C. Zener, Proc. R.Soc. A {\bf 143}, 523 (1934).
\bibitem{oka1} T. Oka, R. Arita, and H. Aoki, Phys. Rev. Lett. {\bf 91}, 066406 
(2003).
\bibitem{oka2} T. Oka and H. Aoki, Phys. Rev. Lett. {\bf 95}, 137601  (2005).
\bibitem{oka3} T. Oka and H. Aoki, Phys. Rev. B {\bf 81}, 033103  (2010).
\bibitem{ecks} M. Eckstein, T. Oka, and P. Werner, Phys. Rev. Lett. {\bf 105}, 146404  (2010).
\bibitem{niu} Q. Niu and M. G. Raizen, Phys. Rev. Lett. {\bf 80}, 3491  (1998).
\bibitem{oka4} T. Oka, arXiv11056.3143.
\bibitem{pb} for a review see e.g. P. Prelov\v sek and J. Bon\v ca, arXiv1111.5931.
\bibitem{lieb} E.H. Lieb and F.Y. Wu, Phys. Rev. Lett. {\bf 21}, 192  (1968).
\bibitem{ovchinnikov} A.A. Ovchinnikov, Zh. Eksp. Teor. Fiz 57, 2137 (1969).
\bibitem{dykh} A.M. Dykhne, Sov. Phys. JETP {\bf 14}, 941  (1962).
\bibitem{davi} J.P. Davis and P. Pechukas, J. Chem. Phys. {\bf 64}, 3129  (1976).

\end{thebibliography}
\end{document}